\documentclass{article}
\usepackage[a4paper, total={6in, 8in}]{geometry}
\usepackage{amsmath,amssymb,amsthm}
\usepackage{algpseudocode,algorithm,graphicx}
\pdfoutput=1 
\newtheorem {theorem}{Theorem}[section]
\newtheorem {lemma}[theorem]{Lemma}
\newtheorem* {defini}{Definition}
\newtheorem*{theorem*}{Theorem}

\usepackage[pdfborder={0 1 1}]{hyperref}

\title{Separating Colored Points with Minimum Number of Rectangles}

\author{Navid Assadian\thanks{Department of Computer Science, University of Victoria {\tt navid@uvic.ca}}
        \and
        Sima Hajiaghaei Shanjani\thanks{Department of Computer Science, University of Victoria {\tt sima@uvic.ca}} 
	\and
	Alireza Zarei\thanks{Department of Mathematical Science, Sharif University of Technology {\tt zarei@sharif.edu}} 
}
 \date{2016}

\begin{document}
\thispagestyle{empty}
\maketitle

\begin{abstract}
In this paper we study the following problem: Given $k$ disjoint sets of points, $P_1, \ldots , P_k$ on the plane, find a minimum cardinality set $\mathcal{T}$ of arbitrary rectangles such that each rectangle contains points of just one set $P_i$ but not the others. We prove the NP-hardness of this problem.
\end{abstract}

\section{Introduction}

Let $P = \{ P_1, \ldots , P_k \}$ be sets of disjoint points on the plain with $n_i=|P_i|$. We say that a rectangle $t$ is $P_i$-rectangle if $t$ contains points from $P_i$ and there is no point of $P_j , j \neq i$, that lies on the $t$.

The separation problem on $P$ with arbitrary rectangles can be defined as follows:

 \begin{defini}[ Minimum Rectangular Separation (MRS):]

\textit{Given $P = \{ P_1, \ldots , P_k \}$, find a minimum cardinality set $\mathcal{T}$ of $P_i$-rectangles such that every point in $P$ is contained in at least one rectangle of $\mathcal {T}$. }
\end{defini}

Bereg et. al. showed that the \textit{Boxes Class Cover} problem, a version of MRS, where the rectangles are axis-aligned is NP-hard \cite{Bereg2012294}. They also provide approximation algorithms for some special cases of the problem, for example when the rectangles are axis-aligned squares.

We show the MRS problem is NP-hard for $k=2$. This implies that MRS is NP-hard for any $k \geq 2$. We call this version of the problem, MRS for $k=2$, \textit{Arbitrary Boxes Class Cover problem (ABCC)}. As this can bee seen as a variant of BCC where rectangles can be in any direction.

Bereg et. al showed that BCC-problem is NP-hard by a reduction from the \textit {the Rectilinear Polygon Covering problem}~\cite{Bereg2012294}. This reduction cannot be used to prove the NP-hardness of ABCC. Here we show the NP-hardness of ABCC by a reduction from newly defined version of the 3-SAT problem, the \textit{NAS-SAT} problem.

\subsection{Outline}
 In section 2, we define a new version of the 3-SAT problem (the NAS-SAT problem), and we prove this problems is NP-hard.

In Section 3, we prove the NP-hardness of BCC by a reduction from the NAS-SAT problem. Then we show how we can modify this and show the NP-hardness of ABCC-problem. 

\section{NAS-SAT}

\textbf {Not All the Same Satisfication problem (NAS-SAT problem):}

We say that a boolean formula $\phi$ is in the form of NAS-CNF if it has the following properties:

\begin{enumerate}
\item
$\phi$ is in CNF form.
\item
All of its clauses has at most 3 literals.
\item
All the clauses with more than one literal have at least one literal in negated form and one literal in non-negated form. 
\item
Any literal appears in at most one clause of size 3.
\end{enumerate}

The satisfication problem on NAS-CNF formulas is called NAS-SAT.

\begin{theorem}
NAS-SAT is NP-Complete.
\end{theorem}

\begin{proof}
NAS-SAT is in NP, as for any given truth assignment the value of the formula cna be computed in polynomial time.

To show the NP-hardness of the problem, we reduce 3-SAT problem to NAS-SAT. 

Suppose that an instance $\phi$ of 3-SAT is given. By the following transformations, we transform $\phi$ in polynomial time to $\phi_{NAS}$ which is an instance of NAS-SAT.

\begin{enumerate}
\item
For every clause $c_i \in \phi$ in form of $c_i = (x \vee y \vee z)$ add the following clauses to $\phi_{NAS}$:
\begin {align*}
& (x \vee y \vee z)  \equiv  (x \vee y \vee {\bar{w}}_i) \wedge (w_i \vee z) \wedge (\bar{w}_i \vee \bar{z}) \\ &\equiv (x \vee y \vee {\bar{w}}_i) \wedge (w_i \vee z \vee False) \wedge ({\bar{w}}_i \vee \bar{z} \vee False) \\ &\equiv (x \vee y \vee {\bar{w}}_i) \wedge (w_i \vee z \vee \bar{T}) \wedge (T) \wedge ({\bar{w}}_i \vee \bar{z} \vee F) \wedge (\bar{F}) 
\end {align*}

\item
For every clause  $c_i \in \phi$ in form of $c_i = (\bar{x} \vee \bar{y} \vee \bar {z})$ add the following clauses to $\phi_{NAS}$:
\begin {align*}
 & (\bar{x} \vee \bar{y} \vee \bar{z})  \equiv (\bar{x} \vee \bar{y} \vee w_j) \wedge (w_j \vee z) \wedge (\bar{w}_j \vee \bar{z}) \\ &\equiv (\bar{x} \vee \bar{y} \vee w_j) \wedge (w_j \vee z \vee False) \wedge ({\bar{w}}_j \vee \bar{z} \vee False) \\ &\equiv (\bar{x} \vee \bar{y} \vee w_j) \wedge (w_j \vee z \vee \bar{T}) \wedge (T) \wedge ({\bar{w}}_j \vee \bar{z} \vee F) \wedge (\bar{F}) 
\end {align*}

\item
Add other clauses to $\phi_{NAS}$.

\item
For every variable in $\phi_{NAS}$ that is appeared in more than one clause of size three, we replace them by the following instruction:

Suppose that the variable $x_i$ is appeared in $t$ different clauses. Therefore we replace them with new variables $x_{i_1}, \dots , x_{i_t}$. For preserving the equivalence of these variables, for every $1<k \leq t$, we insert the following clauses: $$ (x_{i_1} \vee \bar{x_{i_k}}) \wedge (\bar{x_{i_1}} \vee x_{i_k}) $$  

\end{enumerate} 
 $\phi_{NAS}$ is a CNF formula and every clause has at most 3 literals. By the first and second transformations, the third condition of the NAS-SAT problem is satisfied. By the last transformation, the forth condition is also satisfied. Thus,  $\phi_{NAS}$ is an instance of NAS-SAT.

This transformation is done in polynomial time. Therefore, for every instance of the 3-SAT problem, we can construct an instance of NAS-SAT problem in polynomial time which the satisfibility of both of them are equivalent. 3-SAT is NP-hard \cite{Karp1972}. Thus, the NAS-SAT problem is NP-complete. 

\end{proof}

\section{ABCC}

In the rest of this paper, we assume that $B:= P_1$ is the set of blue points, $R:=P_2$ is the set of red points, and $n := n_1 = |B|$. Therefore in ABCC the goal is to cover all the blue points and no red point with minimum number of rectangles.

First we show BCC is NP-hard, then we modify the reduction for ABCC problem.

\subsection{BCC}

\begin{figure}
\centering
	\includegraphics[width= 7cm]{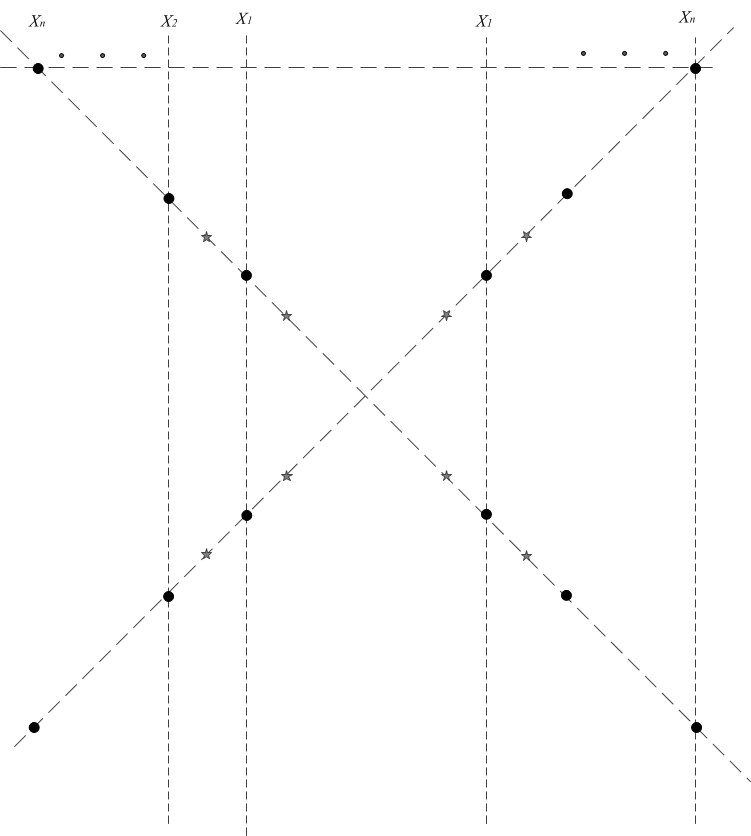}
	\caption{\textit{Variable points} for all the variables. Circles indicate blue points, and stars indicate red points.}
	\label{fi:2}
\end{figure}

Suppose that we have an instance of the NAS-SAT problem, a boolean formula $\phi$. For every variable, we add some points as shown in the figure~\ref {fi:2}. In the figures, circles indicate blue points, and stars indicate red points. We call the points added in this step, \textit {Variable Points}.

In this structure, points from one variable with points from another variable cannot be covered with a same axis-aligned rectangle.
For the points in figure~\ref{fi:2}, we have $2^{\frac{n}{4}}$ different optimal solutions in the BCC problem As in an optimal solution the variable points of each variable can be covered either with the two horizontal or the two vertical rectangles covering these points. We want to use this idea of having exponential possible solutions to reduce NAS-SAT to BCC.

 For every clause, depending on the form of the clause we add different points. The position of these points are described below.

\begin {itemize}
\item
For every clause of size one in the form of $c_i = x_i$, we add a blue point exactly in the middle of the segment that connects the two leftmost variable points of $x_i$, and we add two red points as shown in the Figure~\ref {fi:4-5}. We call the points added in this step \textit{equivalent points} for $c_i$.

\begin{figure}
\centering
	\includegraphics[width=7cm]{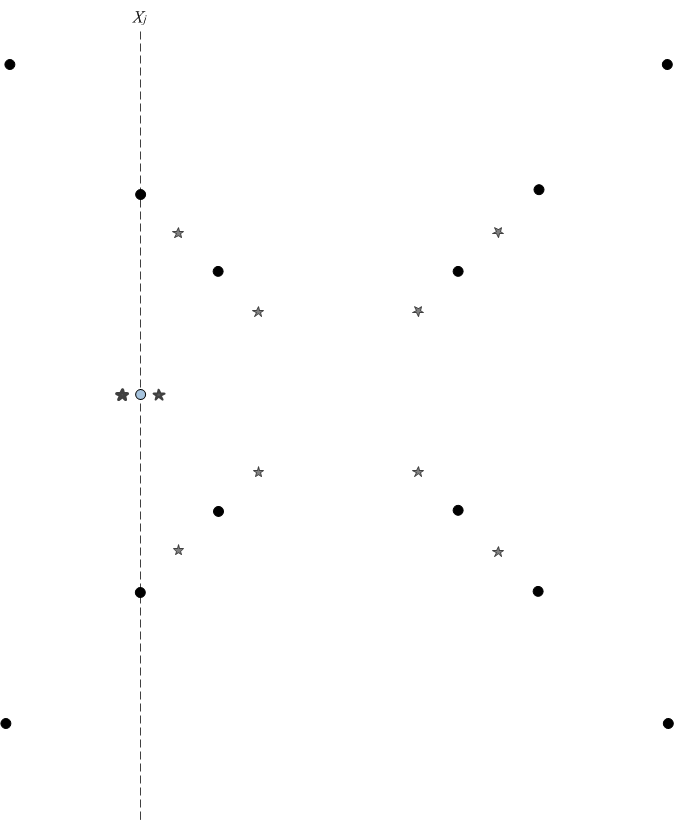}
	\caption{The \textit{equivalence points} for $c_i = x_i$}
	\label{fi:4-5}
\end{figure}

\item
For every clause of size in the form of $c_i = \bar{x_i}$, we add a blue point exactly in the middle of the segment that connects the two lower variable points of $x_i$, and we add two red points on its sides as shown in the Figure~\ref {fi:4-6}. We call the points added in this step \textit{equivalent points} for $c_i$.

\begin{figure}
\centering
	\includegraphics[width=7cm]{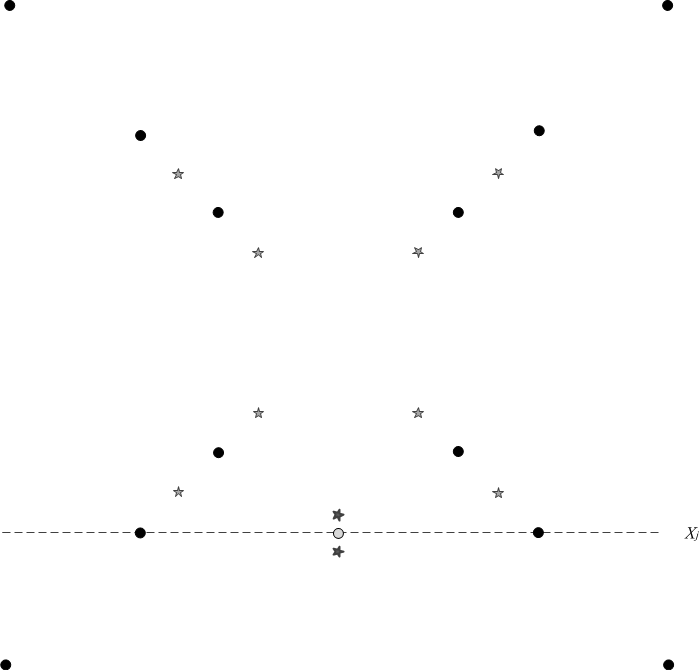}
	\caption{The \textit{equivalence points} for $c_i =\bar{ x_i}$}
	\label{fi:4-6}
\end{figure}

\item
For every clause of size two in the form of  $c_i = x_j \vee \bar{x_l}$, we add a blue point in the intersection of a vertical line that connects the two leftmost variable points of $x_j$ and a horizontal line that connects the two lower variable points of $x_l$. Then, we add four red points around it as shown in Figure~\ref {fi:4-7}. These points are called \textit{equivalence points} for $c_i$.

\begin{figure}
\centering
	\includegraphics[width=7cm]{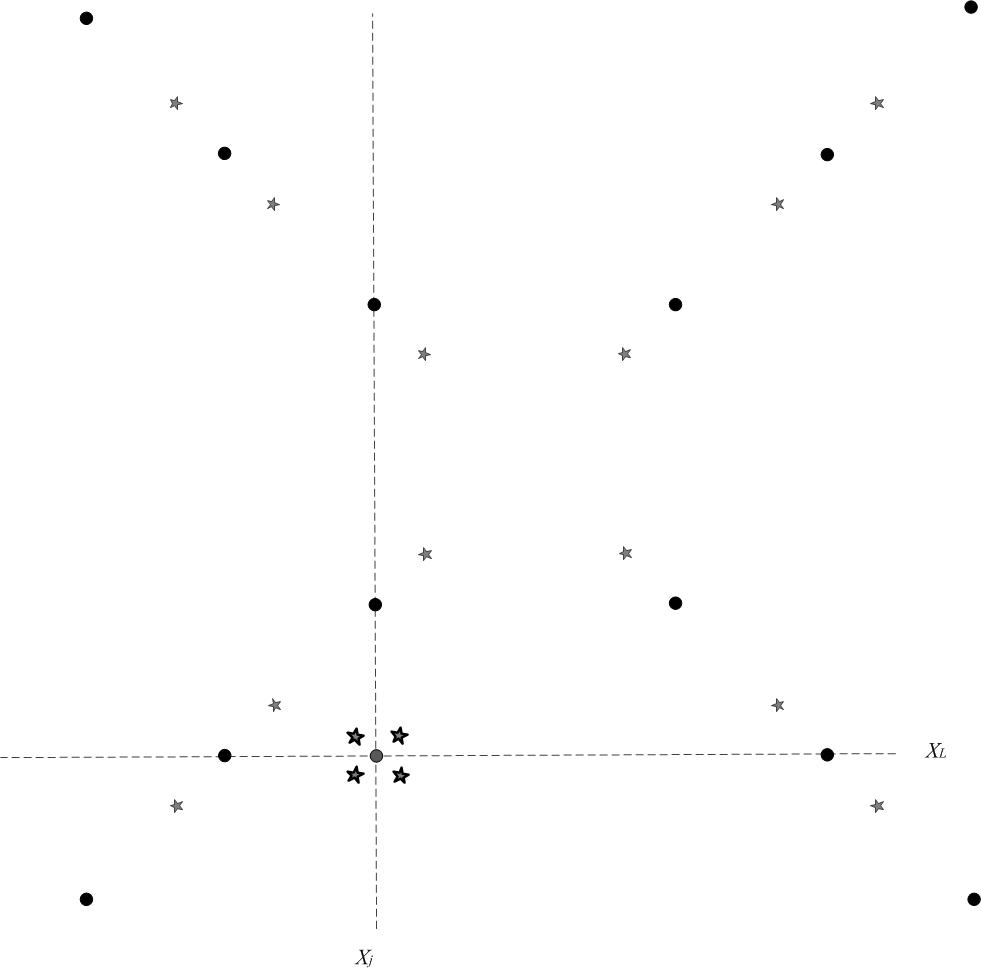}
	\caption{The \textit{equivalence points} for $c_i = x_j \vee \bar{ x_l} $}
	\label{fi:4-7}
\end{figure}

\item
For every clause of size three in the form of $c_i = x_j \vee \bar{x_k} \vee \bar{x_l}$, we add two blue points as follows:

Assume the two vertical rectangles (strips) that cover variable points of $x_j$ and the two upper horizontal rectangles that cover the variable points of $x_k$ and $x_l$. We add a blue point in the intersection region between the leftmost rectangle of $x_j$ and the upper horizontal rectangle of $x_k$. We also add a point in the intersection region between the rightmost rectangle of $x_j$ and the upper horizontal rectangle of $x_l$. The position of these points are shown in Figure \ref{fi:8}. We call these points the \textit{clause points} for $c_i$. 

Then, assume the region which is defined by the intersection of the upper rectangle of the red points which are between the upper rectangles of $x_l$ and $x_k$, and the leftmost rectangle of the red points which are placed on the right side of the leftmost rectangle of $x_j$. We add a blue point in the upper left corner and a red point in the lower right corner of this region. The position of these points are shown in Figure~\ref {fi:8}. We call these points the \textit{helping points}. 

Note that, a single blue-rectangle cannot cover the blue helping point and clause points together.

\begin{figure}
\centering
	\includegraphics[width=7cm]{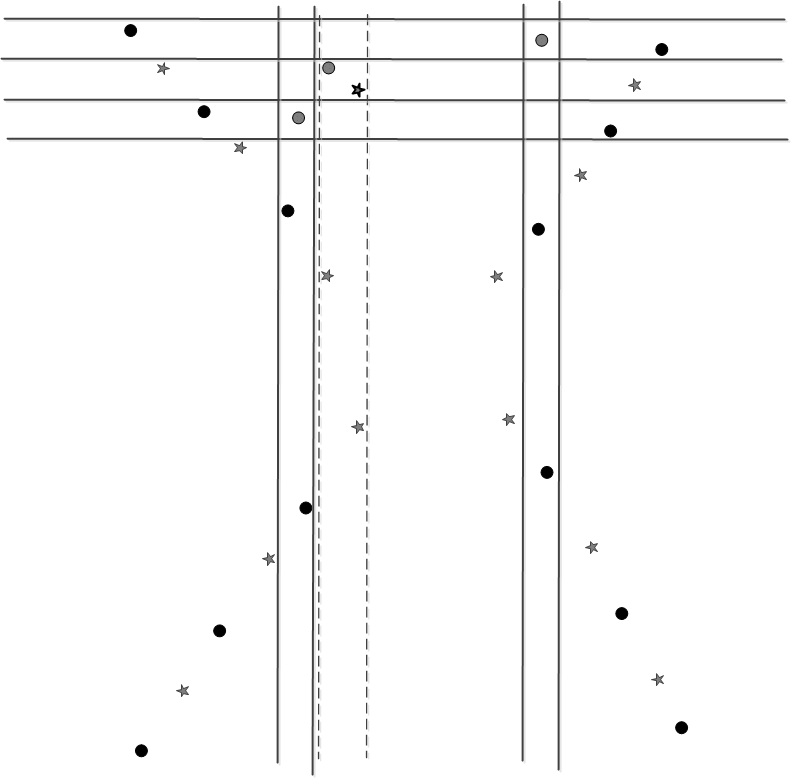}
	\caption{The \textit{clause points} and \textit{helping points} for $c_i = x_j \vee \bar{x_k} \vee \bar{x_l}$}
	\label{fi:8}
\end{figure}

\item
For every clause of size 3 in the form of $c_i = \bar{x_j} \vee x_k \vee x_l$, We add points to the similar places described in the previous part, but with a $\pi/2$ rotation as shown Figure~\ref {fi:9}.

\begin{figure}
\centering
	\includegraphics[width=7cm]{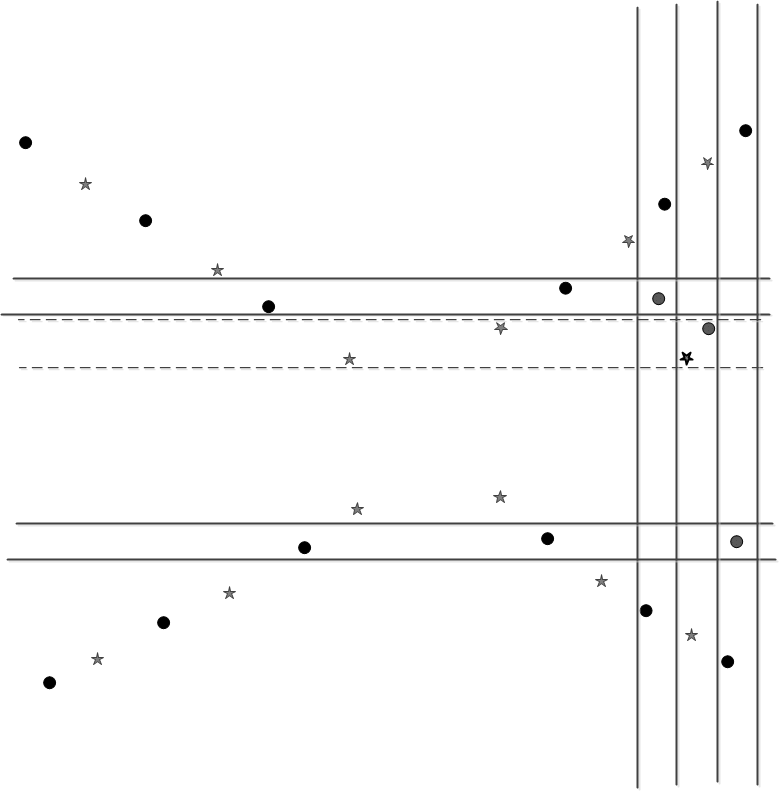}
	\caption{The \textit{clause points} and \textit{helping points} for $c_i = \bar{x_j} \vee x_k \vee x_l$}
	\label{fi:9}
\end{figure}
  
\end {itemize}

To convert any solution for BCC to a solution for NAS-SAT, for any variable $x_i$, if the blue \textit{variable points} of $x_i$ are covered by the vertical rectangles in the solution for BCC, then we assign $x_i$ to `true' in the truth assignment for $\phi$. If the blue \textit{variable points} of $x_i$ are covered by the horizontal rectangles, then we assign $x_i$ to `false' in the truth assignment for $\phi$.

To convert any solution for NAS-SAT to a solution for BCC, for any variable $x_i$, if $x_i$ is assigned to `true' in $\phi$, then we add two vertical rectangles to the solution for BCC to cover all the blue \textit{variable points} for $x_i$. If $x_i$ is assigned to `false' in $\phi$, then we add two horizontal rectangles to the solution for BCC to cover all the blue \textit{variable points }for $x_i$.

\begin{lemma}
\label{l:reduction}
All the blue points can be covered by $2n+m$ blue-rectangles if and only if $\phi$ is satisfiable, where $n$ is the number of variables and $m$ is number of clauses of size three in $\Phi$.
\end{lemma} 

\begin{proof}
For proving this claim, we have to prove following propositions:
\begin {enumerate}
\item
For covering all the blue points, at least $2n+m$ blue-rectangles is needed.
\item
If the blue points are covered by exactly $2n+m$ blue-rectangles in an optimal solution for BCC, then $\phi$ is satisfiable.
\item
If the blue points are covered by more than $2n+m$ blue-rectangles in any optimal solution for BCC, then $\phi$ is not satisfiable.

\end{enumerate}

Proof of 1. To cover all the blue points in this instance of BCC, we need at least two rectangles for each variable. This is because no blue-rectangle can cover the blue \textit{variable points} of two different variables together. To any blue \textit{helping points} we need at least one rectangle, as no blue-rectangle can cover two of these \textit{helping points} or a blue \textit{helping point} and a blue \textit{variable point}. There are $n$ variables and $m$ helping points.  Thus, covering all the blue points needs at least $2n+m$ rectangles.

Proof of 2. By the previous part we know that $2n+m$ rectangles are needed.  If the blue points in BCC are all covered by $2n+m$ rectangles, this means no extra rectangle was used to cover the \textit{clause points} and \textit{equivalence points}. This means the two vertical or horizontal rectangles for each variable are compatible with the form of the clauses, and when we convert this solution with $2n+m$ rectangles to a truth assignment for $\phi$ all the clauses of $\phi$ are satisfied by this truth assignment.

Proof of 3. Assume $\phi$ is satisfiable, then we could convert this truth assignment for $\Phi$ to a solution for BCC. This solution for BCC cannot have more than $2n+m$ rectangles. This is because all the \textit{variable points} are covered by either the two horizontal or the two vertical rectangles. All the blue \textit{clause points} are covered as the clauses are satisfied by this assignment. All th \textit{equivalence points} are covered, as or any \textit{equivalent clauses} at least on of its literals is true in $\phi$. This means the rectangles for the corresponding literal covers the blue point of that \textit{equivalence clause}. Thus, only $m$ more rectangles are needed to cover all the blue points. This is a contradiction as we assumed that the any optimal solution for this instance of BCC needs more than $2n+m$ rectangles.

\end{proof}

By Lemma \ref{l:reduction} this transformation is a reduction, and the transformation is polynomial time. Thus BCC is NP-hard.

\subsection{Hardness of ABCC}
The idea of the reduction from NAS-SAT to ABCC is similar to the reduction from NAS-SAT to BCC. We add the same points described in the previous section for BCC, then we add more red points to the point set in a way that the rectangles in a solution for this instance of ABCC be all axis-aligned rectangles.

For every blue equivalence point, we add red points around it in any direction that there is a blue point in that direction.

For every blue variable point or blue clause point, we add a red point in the intersection between the line that connects it to another blue point and the nearest red strip but not in any blue strip. Note that such region always exists.

We add $O(n^2)$ new red points in polynomial time. In this structure no arbitrary rectangle can cover two blue points that cannot be covered in the reduction from NAS-SAT to BCC. Thus, the size of the solution for this structure is equal to the size of the solution for BCC in the previous stucture.
 Therefore, ABCC problem is NP-hard.



\small 
\bibliographystyle{abbrv}

\end{document}